# An Electronic Textile Embedded Smart Cementitious Composite

Muhammad S Irfan[1], Muhammad A Ali[1], Kamran A Khan[1*], Rehan Umer[1], Antonios Kanellopoulos[2], Yarjan A Samad[3*]

1. Department of Aerospace Engineering, Khalifa University of Science and Technology, Abu Dhabi, United Arab Emirates
2. Centre for Engineering Research, School of Engineering & Computer Science, University of Hertfordshire, Hatfield AL10 9AB, UK
3. Cambridge Graphene Centre, Engineering Department, University of Cambridge, 9, JJ Thomson Avenue, Cambridge, CB3 0FA

*yy418@cam.ac.uk, kamran.khan@ku.ac.ae

## Abstract

Structural health monitoring (SHM) using self-sensing cement-based materials has been reported before, where nano-fillers have been incorporated in cementitious matrices as functional sensing elements. A percolation threshold is always required in order for conductive nano-fillers modified concrete to be useful for SHM. Nonetheless, the best pressure/strain sensitivity results achieved for any self-sensing cementitious matrix are <0.01 MPa$^{-1}$. In this work, we introduce novel reduced graphene oxide (RGO) based electronic textile (e-textile) embedded in plain and polymer-binder-modified cementitious matrix for SHM applications. As a proof of concept, it was demonstrated that these coated fabric-based sensors can be successfully embedded within the cement-based structures, which are independent of any percolation threshold due to the interconnected fabric inside the host matrix. The piezo-resistive response was measured by applying direct and cyclic compressive loads (0.1 to 3.9 MPa). A pressure sensitivity of 1.5 MPa$^{-1}$ and an ultra-high gauge factor of 2000 was obtained for the system of the self-sensing cementitious structure with embedded e-textiles. The sensitivity of this new system with embedded e-textile is many orders of magnitude higher than nanoparticle based self-sensing of cementitious composites. The manufactured e-textile sensors showed mechanical stability and functional durability over long-term cyclic compression tests of 1000 cycles.

# 1. Introduction

Civil infrastructure (dams, bridges, tunnels, road network etc.) is the backbone of our societal and economic growth. Concrete and the related cement-based materials are the construction industry's favourites for a variety of reasons: (i) ease and cost of construction compared to alternatives (e.g. steel); (ii) robustness for a variety of exposure scenarios; (iii) ability to construct a large variety of complex geometries and (iv) excellent mechanical performance [1]. These materials have excellent response in compressive loads and typically display quasi-brittle behavior. To be used for infrastructure projects the presence of steel reinforcement is essential. Steel alleviates the quasi-brittle behavior of concrete and the new composite (reinforced concrete) can sustain well both compressive and flexural loads. The exposure of cement-based infrastructure to a vast number of degrading environments throughout their service life cannot be prevented. The integrity of concrete and cement related materials largely depend on their ability to withstand mechanical and environmental weathering. It is the coupled effect of phenomena such as impact, service loading, chloride/$CO_2$ concentration, pressure/thermal differentials, freeze-thaw and sulfates that can cause severe damage, hence reducing functionality [2]. Damage can manifest itself by localised weakening of the material and can progress in the form of microcracks that gradually reduce the integrity of structures and elements. Under continuous mechanical and environmental stress these microdefects coalesce and expand compromising the mechanical properties of materials. In order to monitor the in-service integrity of the cement-based structures, structural health monitoring (SHM) is deemed vital especially for the safety critical components [3]. SHM can provide real-time data about the condition of a structure by using suitable sensors. This will allow for timely intervention in critical situations minimizing considerably the maintenance regimes and extending the service life of an infrastructure asset.

Conventionally, these sensors are either embedded or applied on the external surfaces of the structure [4]. A number of sensing techniques including, foil strain gauge [5], optical fiber [5, 6], piezoelectric ceramic [7, 8] and shape memory alloys [7] have been studied for SHM of cement-based constructions. Such sensors are usually incompatible with cementitious materials and reduce the strength and durability of the structure [9].

A lot of work has been reported on self-sensing cement-based materials based on fillers like carbon fibers (CFs) [10-12], carbon nanotubes (CNTs) [1, 10, 13, 14], hexagonal boron nitride (h BN) [15], graphene oxide (GO) [15-22], carbon nanofibers (CNFs) [23], graphite nanofibers (GNFs) [9], graphene nanoplatelets (GNPs) [24, 25], carbon black [26-28], steel fibers [29-31], nickel powders [32, 33], conducting rubber [34, 35], and MXenes [36] etc. A major challenge in using these fillers is their homogeneous dispersion in the cementitious matrix especially in large volumes [37]. Most of the reported work involves modification of the cementitious matrix at nanoscale and shows improvement or retention of mechanical properties while providing the self-sensing capabilities via piezo-resistive response. Nonetheless, the application of these fillers at large volumes is a technical and economic challenge. After we introduced RGO coated electronic textiles in early 2014 [38] instead of using nanofillers in the bulk material, some researchers have employed such smart textiles as sensor elements in the field of fiber reinforced composites to monitor the processing parameters and in-service strain [39-41]. Recently, we have developed a simple and scalable application of RGO coated substrates for pressure sensing [42]. The method has been applied to wearable items for real-time health and physical performance monitoring. All the applications have demonstrated both repeatability and scalability.

Various polymers such as urea-formaldehyde resin, unsaturated polyester resin, methylmethacrylate, epoxy resin, furan resins, polyurethane resins and waste tire rubber have been used by researchers to create polymer-modified mortar (PMM)/polymer-modified concrete (PMC) [43-47]. There are two major reasons for adding polymeric materials to the cementitious matrix: (i) to re-use the waste polymer; and (ii) to change specific properties of the cement-based materials e.g. density, fatigue life, toughness, brittleness, and moisture absorbance [44]. The effect of polymer modification on the piezo-resistivity of self-sensing polymers has not been reported.

In this work, we present the application of RGO-coated Nylon textile as the piezo-resistive strain-sensing element in a cement-based system. A comparison was also made between matrices with and without polymer modification. It was shown that the RGO-coated piezo-resistive fabric provides a novel route for making self-sensing cement-based materials and the addition of polymer in the matrix improves the sensing and bonding capabilities of the structure to a great extent.

## 2. Materials and methods

### 2.1. Preparation of RGO coated Nylon fabric

The e-textile was prepared by coating graphene oxide (GO) on a commercially available Nylon fabric, supplied by Gurit®, as a substrate. The Nylon has been chosen as an example in this study, however, any textile fabric compatible with GO can be used as demonstrated in our earlier studies (Ali et al. 2017). For the coating process, the pristine fabric was soaked in a GO solution. The GO, supplied in the form of aqueous paste, was obtained from Abalonyx AS, Norway. The aqueous acidic GO paste contains 25 % GO, 74 % water and 1-1.5 % HCl by weight. The GO solution was prepared by diluting aqueous acidic GO paste of concentration about 100 mg/ml to a concentration of 2.5 mg/ml. The solution was sonicated in a bath sonicator at a frequency of 50 Hz for 30 minutes with the bath temperature maintained at 47°C. A sheet of the Nylon ply was soaked in the solution for 24 hours and then dried in a controlled environment at 80 °C for 5 hours. Once the GO was deposited on the samples, it was reduced by heating under the same controlled environment at 170 °C for 24 hours. The coating process is repeated until uniform coating throughout the fabric was achieved. The reduced graphene oxide (RGO) coated fabrics exhibit a smooth coating that wraps every fiber. The appearance of the coated fabrics is like that of a shiny grey fabric (see Fig. 1). Under a scanning electron microscope (SEM), the coating looks very smooth, adhering well to the fibers' surface (see Fig. 1).

## 2.2. Compression Testing Sample Preparation

The prepared e-textiles were subsequently embedded in plain and polymer modified cementitious matrices. For the preparation of the matrices a slurry was prepared using plain white cement obtained from a local supplier (UltraTech®). The cement was thoroughly mixed with water in in a water to cement ratio of 0.67 (water to cement ration of 2:3) to form the slurry. The samples were prepared by pouring the slurry in metallic molds (200x10x5 mm) in two stages. Initially, the slurry was poured to the middle of the mold, then a small strip of the RGO coated fabric was placed on it and subsequently the mold was filled up to the top. The resulting product was a sandwich structure. The molds were then placed in an oven at 50 C and ambient humidity (20%) to accelerate the setting of the cementitious matrix. For the polymer-modified samples, 2% by weight Sil-Poxy™ Silicone Rubber Adhesive (Smooth-On) was mixed with the cement and water. The samples were prepared in the same exact way as described for above. The process is shown schematically in Fig. 1b & S2b.

## 2.3. Compression Testing

An Instron 5969 universal testing machine with a 2 kN static loading capacity, was used for the mechanical compression tests. Each sample was placed at the bottom stationary platen of the testing frame. The top section was an indenter with rectangular cross-section of 50x10 mm. Two electrical connections were taken along the shorter side of the sample by applying copper strips using silver paste (PECLO® conductive silver paint from TedPell®). A linear force profile with an increment of 10 N/s was applied to the samples to a maximum load of 1.95 kN. After reaching the target maximum load, the unloading cycle was initiated. The electrical current was recorded using an electrochemical workstation (Autolab 302 N), and the load vs. deflection curve was recorded using the Instron data acquisition system. The current values were later converted to resistance using Ohm's law and subsequently fractional change in resistance was obtained. A fixed voltage of 1V was applied and the corresponding current was recorded during loading and unloading periods. The experimental set-up is shown schematically in Fig. 1c & S3.

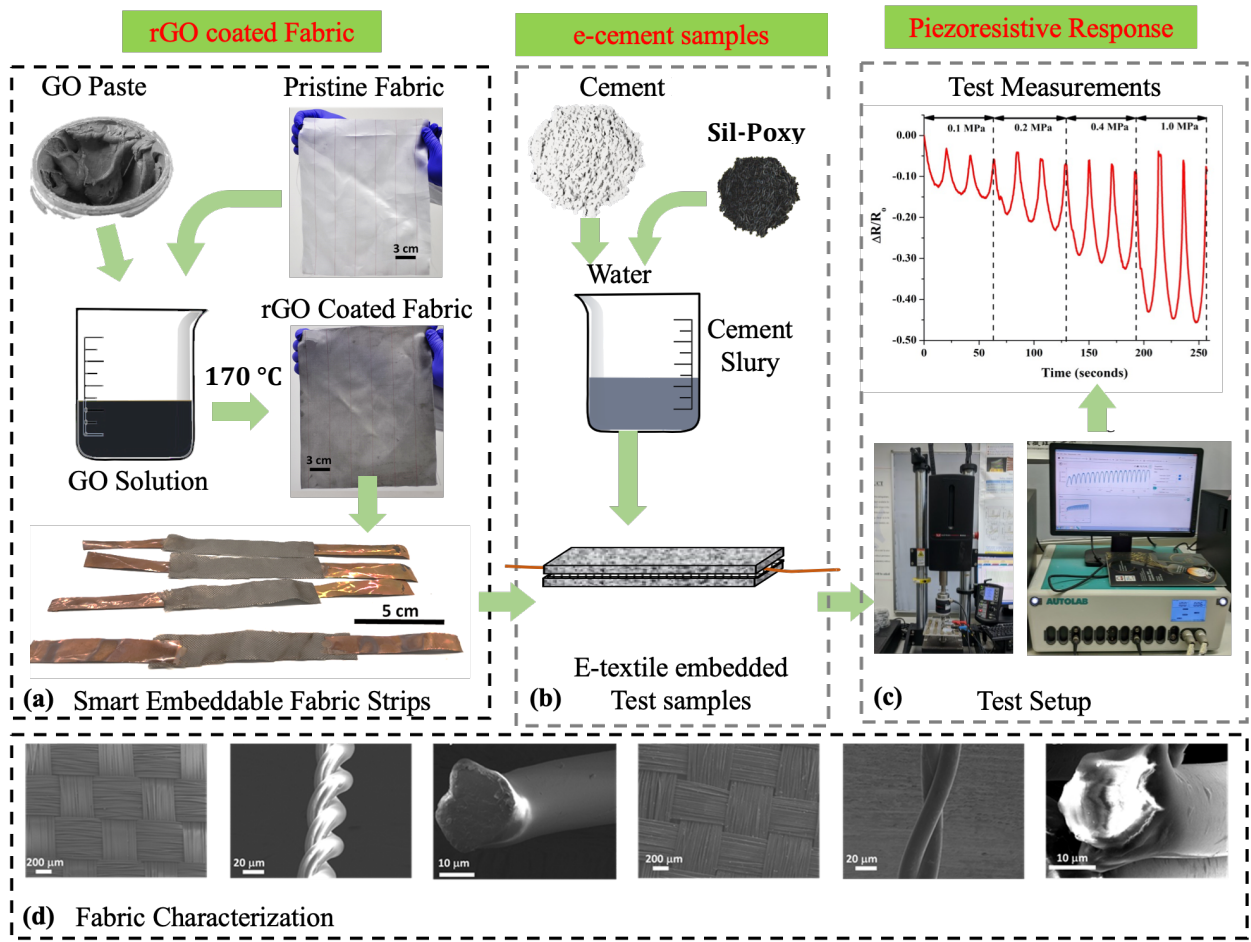

**Fig. 1 Manufacturing of e-textile embedded pure cement and polymer-modified cement structures. a.** Process for coating of Nylon fabric with RGO**. b.** Schematic representation of manufacturing process for e-textile embedded cement structures**, c. Piezoresitive performance measurement setup Testing of e-textile embedded d** SEM images of RGO coated Nylon fabric.

## 3. Results and Discussion

### 3.1 Piezoresistive Response

The e-textile embedded pure and polymer-modified cementitious samples were subjected to a loading-unloading cycle at a maximum compressive stress of 3.9 MPa (1950 N). The fractional change in resistance (FCR) as a function of stress for loading and unloading regime is shown in Figure 2a. An FCR of ~35% is recorded with an application of a compressive stress of 3.9 MPa for the plain sample. For the polymer modified matrix, the FCR is ~60% for the same stress. This is attributed to better transfer of load to the e-textile in the polymer-modified composite. The pressure sensitivity and the corresponding strain sensitivities (gauge factors) for these measurements can be estimated by the following equations [48]:

Pressure sensitivity (PS):

$$PS = \frac{PCR}{Compressive\ Stress\ (MPa)} \quad (1)$$

Gauge factor (GF):

$$GF = \frac{PCR}{Compressive\ Strain\ (\epsilon)} \quad (2)$$

The pressure sensitivity (PS), estimated using equation (1), for a compressive strength of 3.9 MPa for the plain matrix is about 0.1 MPa$^{-1}$ and for that of polymer-modified specimen is 0.15 MPa$^{-1}$. This pressure sensitivity is at least 25 orders of magnitude better than state-of-the-art particulate matter reinforced smart concrete [49]. The maximum compressive strength of 3.9 MPa corresponds to a maximum compressive strain of approximately 0.03% [50, 51]. The gauge factor (GF), as estimated using equation (2), for the plain cementitious matrix is > 1100, which is at least two orders of magnitude and 30 times better than best reports on self-sensing concrete based on conductive particulate matter reinforced cement-based composites [52].

The plain cementitious sandwich structure with embedded e-textile performs better in terms of pressure and strain sensitivity than the polymer-modified one. Nonetheless, the response for the polymer-modified samples is smooth and the curve follows an elliptical locus during the loading and unloading. The corresponding behavior of the non-modified matrix possesses a relatively linear response. Cement is a material rich in minerals such as calcium silicates, aluminates and aluminoferrites. These minerals when in contact with water they undergo a chemical reaction, known as hydration, and with the process of time they transform from a powder form to a network of solid fibrous crystals [53]. The polymer modification of the cementitious matrix impacts the evolution of this cementitious crystalline network. The viscoelastic nature of polymer chains would reduce the brittleness of the bulk structure and would increase the toughness of the bulk material. A schematic representation of the material during loading-unloading cycle is shown in Figure 2b. It is envisaged that the presence of the polymeric chains in the matrix would affect the load transfer to the fabric and hence, the piezo-resistive response.

The FCR under various cyclic compressive loading (0.1 to 1.0 MPa) is shown in Figs. 2c and 2d. Both plain and polymer-modified cementitious samples showed piezo-resistive response to the applied cyclic compressive stresses (0.1 to 1 MPa). The measurements against applied load are shown in Fig. S4. The small compressive strength of 0.1 MPa causes an FCR of ~15% corresponding to a pressure sensitivity of 1.5 MPa$^{-1}$ for the polymer-modified composite. The compressive stress of 0.1 MPa in the plain cementitious matrix causes a change in resistance of ~2% corresponding to a pressure sensitivity of 0.2 MPa$^{-1}$ but a gauge factor of ~2000 as the strain in such low pressures for such matrix is a maximum of 0.001%. This is due to the new system of embedding the e-textile directly inside the cementitious matrix causing a maximum load transfer to the e-textile at smallest overall strain in the host matrix. To an extent the embedded textile operates like a spinal cord within the composite, sensing the stress and feeding the information to a processing unit.

The changes in resistance as a result of the applied load can be explained via two main mechanisms, namely percolation theory and quantum tunneling effect theory. When subjected to compressive strains, the number of RGO-to-RGO contacts increases to form more conductive paths, and the gaps between the RGO particles decrease, leading to the manifestation of the tunneling effect, thereby causing a decrease in electrical resistance leading to the piezo-resistive phenomenon [1, 54]. Also, there is variation in the shape of the peaks of FCR between plain and polymer-modified matrices highlighting possible differences in the way the electrical current is passing through the material. The FCR response as a result of applied compressive stress is shown in Figure 2a. The FCR with applied stress is about linear for the non-modified cementitious matrix and e-textile. The FCR response as a result of applied stress for the polymer-modified cementitious matrix and e-textile structure is non-linear, which could be linked viscoelasticity of polymeric chains.

The functional durability of the sensors was tested by applying a cyclic compressive stress of 0.2 MPa for 1000 cycles as shown in Figs. 3a and 3b. Once the electric polarization was attained, both sensors showed very good functional durability. This also validates the mechanical durability of the sensors as well as the quality of the RGO coating on the Nylon fabric under compression loading. The current-voltage curves of plain cement and polymer modified cement sensors under different compressive loads (from 0 MPa to 1.0 MPa) are shown in Figures 3c and 3d. The linear current-voltage curves within the applied compressive stress range indicate that the composites obey the Ohm's law well, showing the stability of the conductive network. The resistances of the sensors in both cases are showing a decreasing trend tendency with an increase in compressive stress.

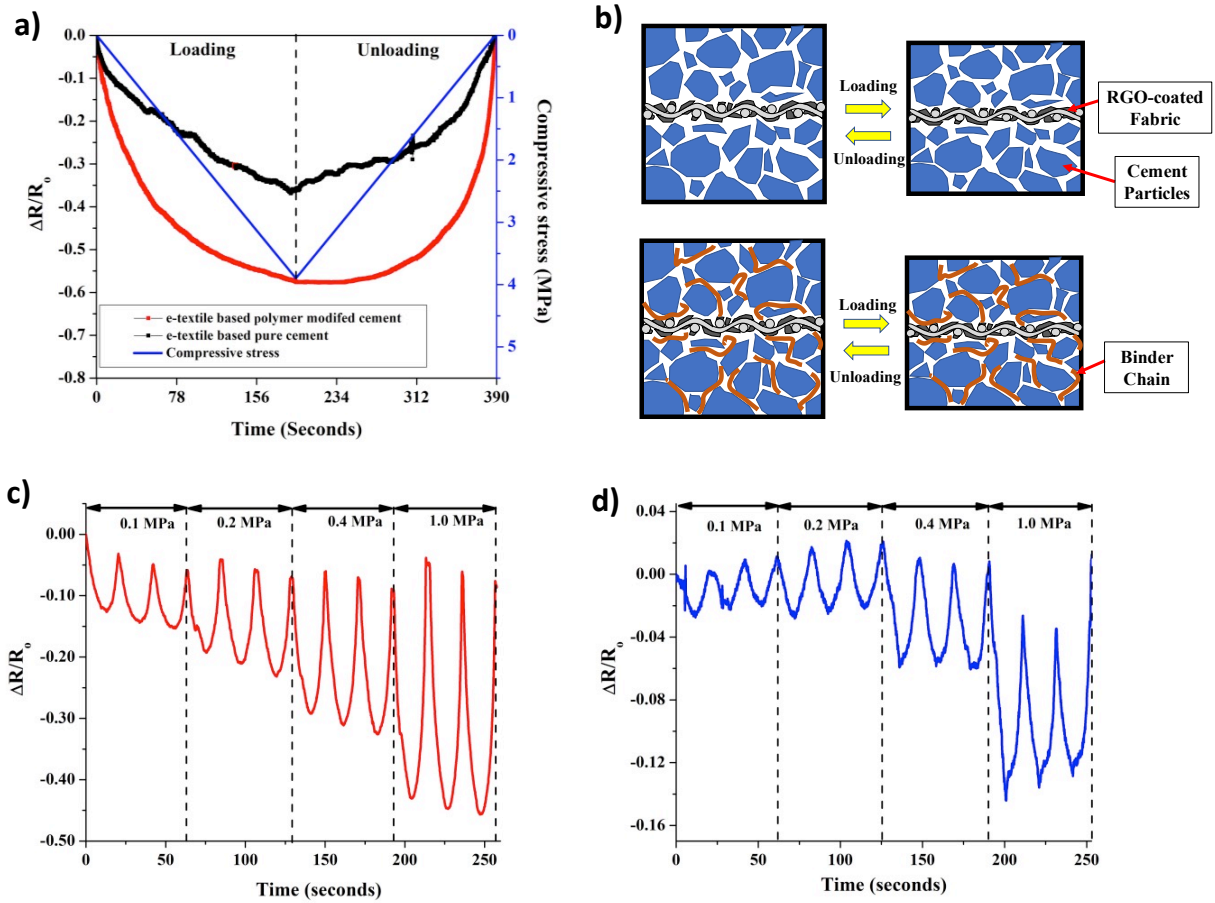

**Fig. 2 FCR response to loading and unloading of e-textile embedded pure cement and polymer-modified cement. a** Comparison of FCR response for e-textile embedded pure cement and polymer-modified cement. **b** Schematic representation of the material during loading-unloading cycle**.** FCR response under various cyclic compressive loading (0.1 to 1.0 MPa): **c** e-textile embedded polymer-modified cement. **d** e-textile embedded pure cement.

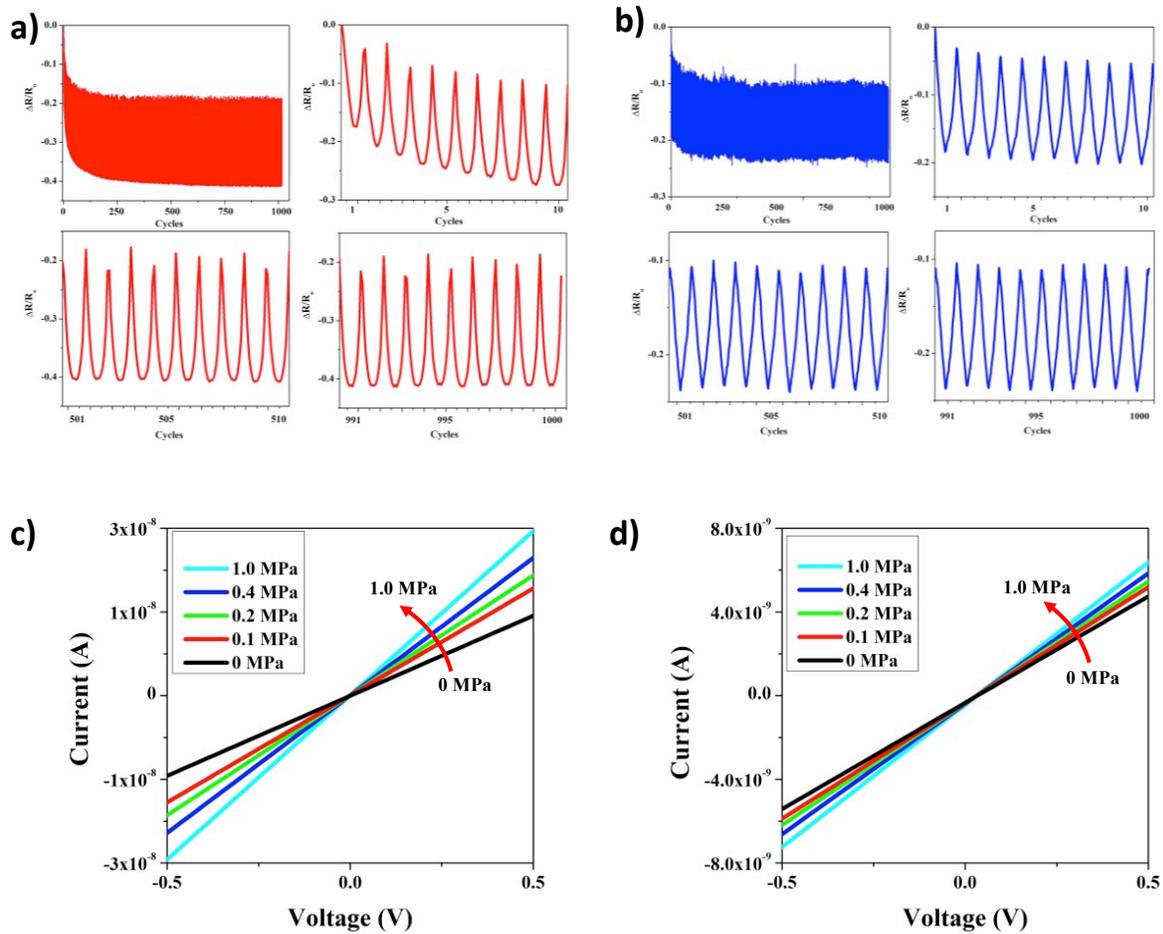

**Fig. 3 Functional durability over time for 1000 cycles at 0.2 MPa compressive stress and Current–voltage (I–V) curves measured on cementitious samples with various compressive stress ranging from 0 to 1 MPa. a** FCR response of e-textile embedded polymer modified matrix over 1000 cycles. **b** FCR response of e-textile embedded pure cementitious matrix 1000 cycles. **c** IV curves for e-textile embedded polymer modified matrix. **d** IV curves for e-textile embedded pure cementitious matrix.

### 3.2 Modeling cyclic response of piezoresistive materials

In this study, we borrowed the concept of circuit theory and employed the analogy of fault current decay to provide the analysis of the mechanisms of cyclic asymmetric resistance decay and symmetric resistance changes during long-term cyclic mechanical loading of piezoresistive materials. Unlike rated resistors, the resistance changes within the cementitious sandwich structure containing the e-fabric under cyclic mechanical loading involve transient components that cannot be ignored when dealing with the piezoresistive materials characterization. Although compared with the steady state resistance component, the transient part of the resistance change is short-lived and lasts few cycles. It needs to be included in the overall resistance change behavior of the piezoresistive materials operated under cyclic loads within a similar time frame when the transient part is still very much active.

The electrical resistance of the e-textile based composite under mechanical cyclic loading can be attributed to two major sources, i.e., intrinsic resistance of the constituents, such as resistance of the cementitious matrix, polymeric binder and RGO coated textile fabric, and contact resistance, which is the resistance between connecting RGO flakes within the textile fabric, the resistance between RGO and the host matrix. The intrinsic resistance of the material depends upon its ability to resist the transmittance of electrons. The intrinsic resistance of graphene depends upon its flakes' geometric parameters, their orientation and surface condition. Resistance to ionic conduction is the major source of resistance in the cement matrix while polymer is usually considered as an insulator.

In this study, we assumed that the bulk resistance of the piezoresistive materials under cyclic mechanical loading is governed by intrinsic resistance of RGO coated fabric and number of graphene flake contacts, area, their packing density, their orientation and deformation mechanisms. It was assumed that the intrinsic resistance contributes to the steady state symmetric resistance change response while the contact resistance shows transient response until it decays completely. As a result, the piezoresistive materials can be consider as higher order systems consisting of several resistors in a complex network, in which all resistor branches (conductive path) contributing to the current, Figure 4 (a). The transient response (resistance or current) under mechanical cyclic loading consists of multiple components, each with a different time constant. The time constant (measured in seconds) is a measure of the steepness of the transient response and it can be different for each component in a composite material system, and may depend on the individual elements' resistances, inductance as well as the frequency of the applied loading. Computing the exact component of the transient resistance will require solving set of differential equation, which is a challenging task. Moreover, if the exact components of the transient resistance are calculated, we would still need to come up preferably with a single exponential term to be able to use the concept of symmetrical based rating for piezoresistive materials.

To handle this problem, here, we used Thévenin's Theorem to simplify our complex electrical circuit to an equivalent two-terminal circuit with just a single constant voltage source in series with two resistances (or impedance) representing a transient resistance and steady state resistance change, respectively, as shown in the Figure 4(b) inset and can be expressed as

$$R(t) = R(t)_{transient} + R(t)_{symmetric} \tag{3}$$

Since, the time-constant is a physical characteristic of the system, so regardless of the complexity of the circuit, the transient resistance is essentially a decaying resistance can be expressed with only one time-constant following circuit reduction techniques such as Thévenin method. What naturally follows is that each material constituent branch may provide a transient resistance with the same time-constant. The transient resistance in such circuit is generally modeled as an exponentially decaying resistance and has the general form of $R_0 \, exp^{(-t/\tau)}$, Where $R_0$ is the initial resistance and $\tau$ is the decay constant. The time constant ($\tau$) depends on the ratios of the material constants of transient to steady state resistance changes that constitutes the piezoresistive materials system under cyclic mechanical loading. Hence, when the applied cyclic load attempts to increase the number of conductive paths, the complete decay of the transient resistance dictate by the values of material constants related to contact resistivity. The larger the parallel resistive network form, the longer it takes for the transient resistance to fade away. The general form of the total resistance change can be expressed as [55]:

$$R(t) = R_0 \left[1 - \sum_i^N R_i \left(1 - exp\left(-\frac{t}{\tau_i}\right)\right)\right] + \left[\frac{a_0}{2} + \sum_i^\infty a_n \cos(n\omega t) + \sum_i^\infty b_n \sin(n\omega t)\right] \quad (4)$$

This resistance change expression is a generalized expression to represent the decay of resistor network with multiple time decay constant subjected to any periodic cyclic loading. The first term represents the resistance decay in a piezoresistive material where multitude decay processes due to decreasing contact resistance give rise to dispersion of the decay time. Where $R_0$ is the initial resistance, $R_i$ is the dimensionless material constant representing the weighted contribution of each parallel resistor branch towards the total resistance decay and $\tau_i$ represents the corresponding time constant defining steepness of the decay process[1]. The second term characterizes the symmetric part of the piezoresistive response due to periodic cyclic loading and can be represented by Fourier series. Where, $a_0, a_n, and\ b_n$ are the coefficients, and $\omega$ is the circular frequency. The $a_0$ is the amplitude of the symmetric steady state resistance change and can depend on amplitude of the cyclic mechanical loading for the case considered[2].

Figure 4(a) shows the schematic of the piezoresistive material showing the initial conductive paths available within the network. With an application of the cyclic load more permanent conductive paths are produced and eventually after few cycles the steady state condition of

---

[1] It is also important to notice that the time constant of the system is solely determined by the circuit's elements resistance; no other variable such as voltage amplitude, cyclic mechanical loading amplitude and phase or the initial conditions of the circuit has any impact on the time-constant.
[2] While the rate of decay of the transient resistance depends only on the circuit elements, its amplitude depends on input voltage, cyclic mechanical loading circuit elements, and, importantly, on the initial conditions of the system.

resistance change under cyclic load is achieved as shown in Figure 4(b). Equation (4) was applied assuming that both the cementitious matrix/rGO coated fabric and cementitious matrix /polymer/rGO coated fabric can be represented by the Thevenin's Method with one decay constant. For periodic triangular cyclic loading, the parameters for symmetric resistance change are: $a_0 = A = Amplitude\ of\ Symmetric\ part, a_n = \sum_i^\infty \frac{-4A}{(2n-1)^2\pi^2} \cos(2n-1)\omega t$ and $b_n = 0$

For transient resistance change the parameters for cement/rGO coated fabric are found to be $\tau_1 = 358\ sec$, $R_1 = 1$ while for cement/binder/rGO coated fabric the parameters are $\tau_1 = 732\ sec$, $R_1 = 1$. Figures 4(c) and (d) show the comparison of the experimental data and the responses generated by the Equation (4) using Thevenin method.

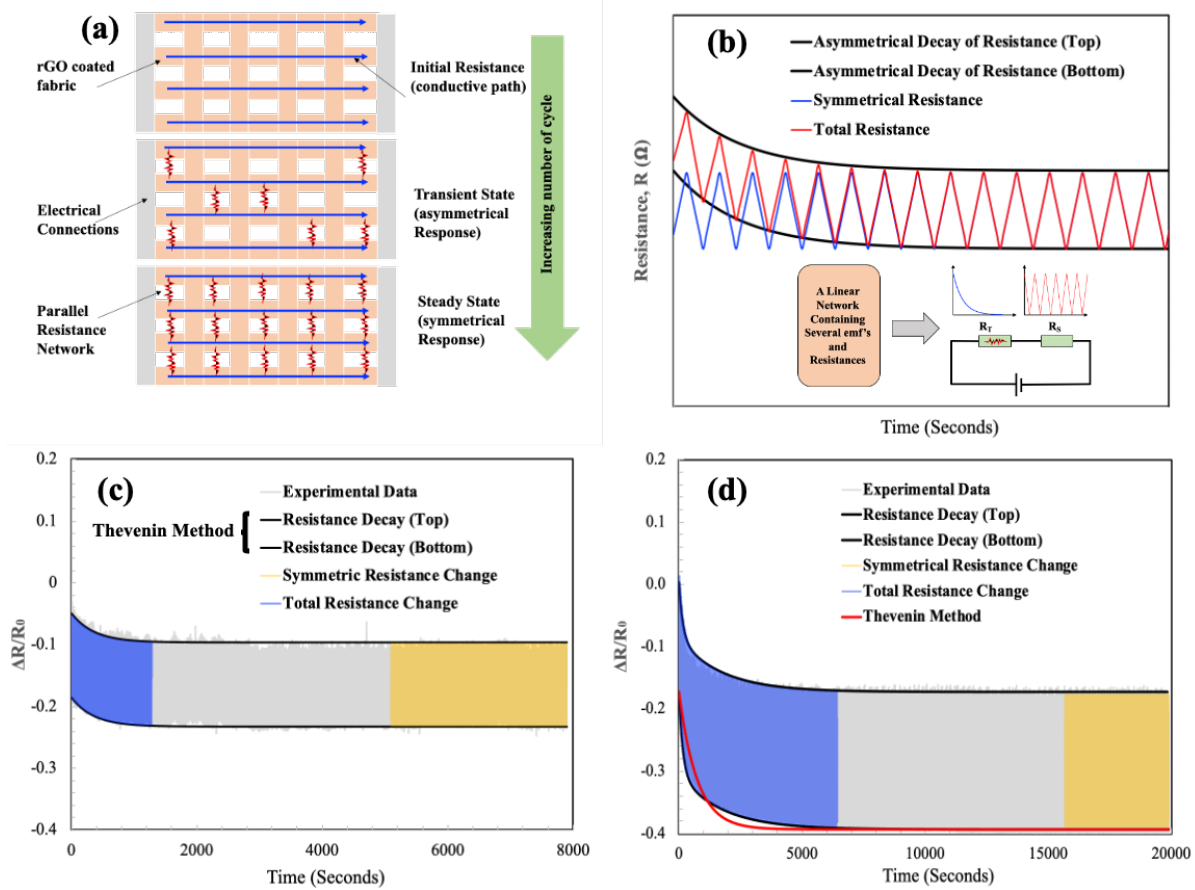

**Fig. 4. Modeling resistance change in a piezoresistive materials due to cyclic mechanical loading. a** Schematic of the mechanism of formation of the new conductive paths during cyclic loading. **b** A simplified two resistance model representing symmetric and asymmetric part of resistance evolution during cyclic loading. **c** Modeling of change in resistance per unit original resistance for e-textile embedded pure cementitious matrix using Thevenin method (black line). **d** Modeling of change in resistance per unit original resistance for e-textile embedded polymer modified matrix using Thevenin method (red line) and higher order method (black line) with multiple decay constants.

For cementitious matrix/rGO coated fabric, the Thevenin's Method found to be a good approximation, however, one time-constant is not enough to represent the resistant change response of cementitious matrix/polymer/rGO coated fabric. It can be seen that the Thevenin's method results (shown in red color line) produce more decay than the actual decay of the piezoresistive material, so we believe that the presence of the polymeric-chains requiring more than one time constant to capture the decay response. Therefore, we used two time-constant model to capture the decay response of cement/binder/rGO coated fabric response and found that due to the presence of the polymeric-chains in the matrix the decay of transient resistance change takes longer time. For multiple decay mechanisms in cement/binder/rGO coated fabric the parameters are found to be $\tau_1 = 160\ sec$, $R_1 = 0.6$, $\tau_2 = 1750\ sec$, $R_2 = 0.4$. The presence of the polymeric binder produces additional time-constant due to the delayed response of the inherited viscoelastic behavior of the polymer under cyclic mechanical loading.

## 4. Conclusion

The use of RGO based e-textile embedded in cementitious matrices has been successfully demonstrated. It was shown that the novel e-textiles possess sensitivities as high as 1.5 MPa$^{-1}$ and gauge factors as high as 2000, several orders of magnitude better than any other state-of-the-art method of self-sensing used in cement-based composites. Furthermore, the sensors showed very good mechanical stability and functional durability over long-term tests of 1000 cycles. We presented the analysis of the mechanisms of cyclic asymmetric resistance decay and symmetric resistance changes during long-term cyclic mechanical loading of piezoresistive materials. A simple relationship was developed and showed that the plain matrix resistance decay faster as compared to the polymer modified cement. For cementitious matrix/rGO coated fabric, the Thevenin's Method found to be a good approximation of an equivalent circuit, however, non-Thevenin's calculations with at least two different decay time constant are needed to represent the resistant change response of cementitious matrix/polymer/rGO coated fabric as a direct

# Acknowledgements

This publication is based on work supported by Abu Dhabi Award for Research Excellence (AARE-2019) under project number 8434000349 and by the Khalifa University of Science and Technology internal grants CIRA-2018-15, and FSU-2019-08. Dr Yarjan acknowledges his research visit support of Khalifa University of Science and Technology under Research Publication Award (Khan) with Project No. 8474000195.

# Supporting Information

# A Cementitious Composite with Central Nervous System

**Muhammad S Irfan[1], Muhammad A Ali[1], Kamran A Khan[1*], Rehan Umer[1], Antonios Kanellopoulos[2], Yarjan A Samad[3*]**

5. Department of Aerospace Engineering, Khalifa University of Science and Technology, Abu Dhabi, United Arab Emirates
6. Centre for Engineering Research, School of Engineering & Computer Science, University of Hertfordshire, Hatfield AL10 9AB, UK
7. Cambridge Graphene Centre, Engineering Department, University of Cambridge, 9, JJ Thomson Avenue, Cambridge, CB3 0FA

*yy418@cam.ac.uk, kamran.khan@ku.ac.ae

Raman spectra for GO and partially reduced RGO are shown in Figure S1a. Two fundamental vibrations can be observed in the range of 500 and 2500 cm$^{-1}$. From Lorentzian fitting of the curves it can be seen that the D vibration band, which is formed from a breathing mode of j-point photons of $A_{1g}$ symmetry is at 1356 and 1358 cm$^{-1}$ for GO and RGO, respectively. G vibration band appeared at 1592 for GO and 1589 cm$^{-1}$ for RGO from first-order scattering of $E_{2g}$ phonons by sp$^2$ carbon. The D and G bands in the Raman spectrum depict the disorder bands and the tangential bands, respectively [i,ii]. The roughness measure (Figure S1b) from atomic force microscopy topographical image (Figure S1c) shows a thickness of about 3nm corresponding to about 4-5 layers of GO sheets with the edges folded for up to 8nm. The folding at the edges is due to the large size of the sheets.

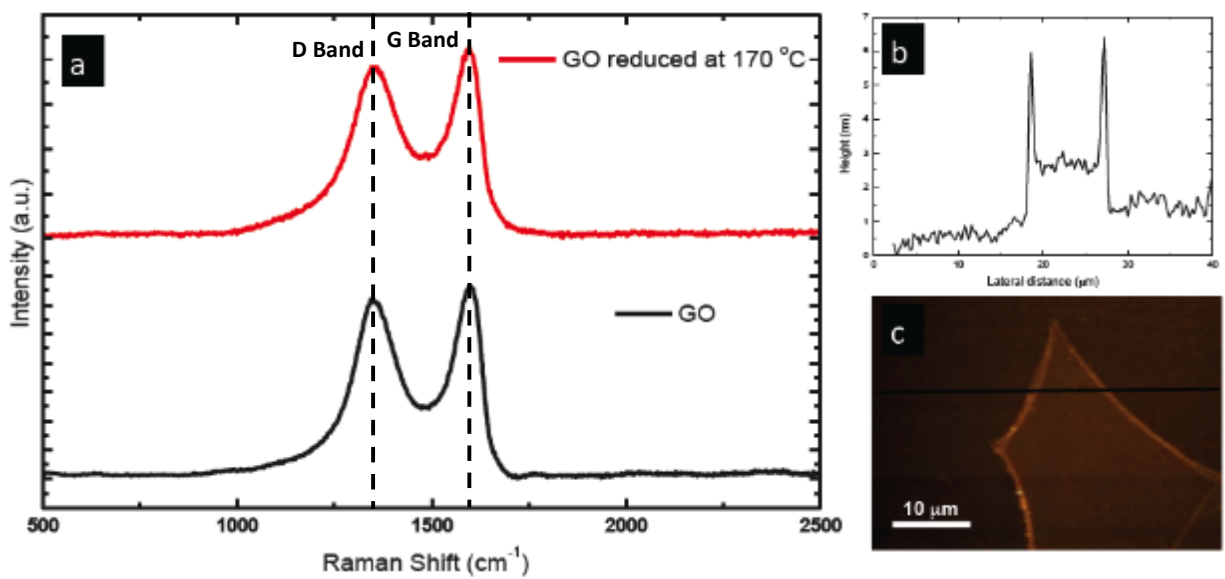

Figure S1: Raman spectra of graphene oxide (GO) coated and partially reduced graphene oxide (RGO) coated fabric

**Fabrication process of RGO coated e textile**: For the fabrication of RGO coated e-textile, the graphene oxide (GO) solution was prepared by using aqueous GO paste (Abalonyx AS, Norway). The aqueous acidic GO paste contains 25 % GO, 74 % water and 1-1.5 % HCl by weight. The GO solution with a concentration of 2.5 mg/ml was prepared by diluting aqueous acidic GO paste to a concentration of 10 % by weight which corresponds to a GO. The solution was sonicated in a bath sonicator at a frequency of 50 Hz for 30 minutes with the bath temperature maintained at 47$^o$C. A sheet of the Nylon fabric (Gurit®) was soaked in the solution for 24 hours and then dried in a controlled environment at 80 °C for 5 hours. Once the GO was deposited on the samples, it was thermally reduced by heating the textile fabric at 170 °C for 24 hours in an oven. The process is shown schematically in Fig. S2a.

**Manufacturing of e-textile embedded cement structure:** For manufacturing the e-textile embedded cement samples for compression testing, first the strips were cut from RGO coated e textile and electrical connections were made on edges of the strips by attaching copper strips using silver paste (Product 16062, PELCO® Conductive Silver Paint, TedPella). Once the e-textile strips were prepared a slurry of cement and water as prepared by mixing. Initially, the slurry was poured in the mold (200×10 mm) and a small strip of the RGO coated fabric was placed on the wet cement mixture. Additional slurry was poured on the top surface to form a sandwich structure. The mold was then placed in an oven for fast curing of the cement. In the second set of samples 2% by weight Sil-Poxy™ Silicone Rubber Adhesive (Smooth-On) was mixed with the cement and the samples were casted in the same way as described for the plain cement. The process is shown schematically in Fig. S2b.

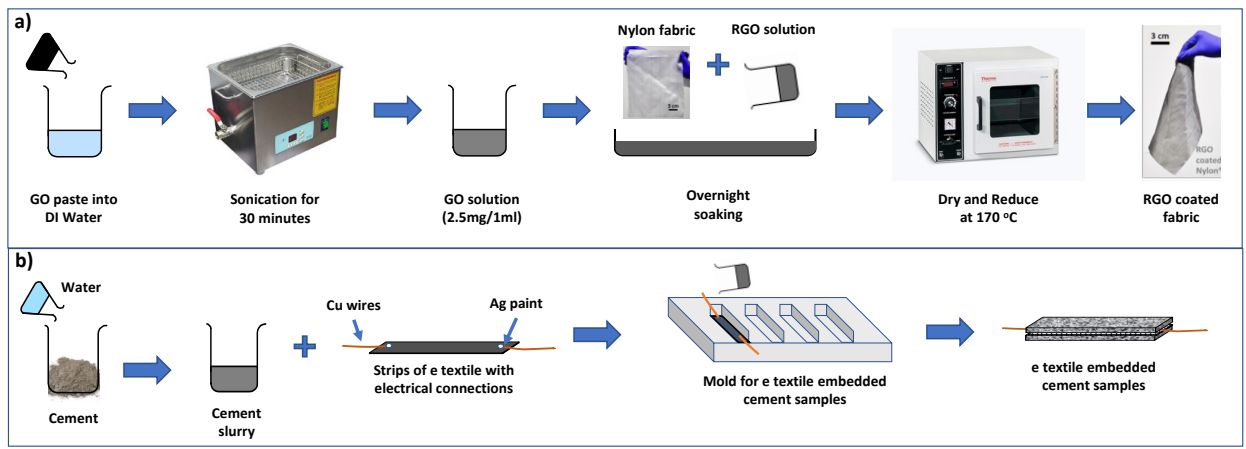

Figure S2 (a) Fabrication process of RGO coated e textile. (b) Manufacturing of e-textile embedded cement structure.

Piezo-resistive response of the e-textile embedded cement samples was obtained by applying the compression loading using an Instron 5969 universal testing machine (Figure S3a) with a 2 kN static loading capacity and recording current using Autolab 302 N electrochemical workstation (Figure S3b). A fixed voltage of 1V was applied during the experiments. The current values were later converted to resistance using Ohm's law and subsequently fractional change in resistance was obtained.

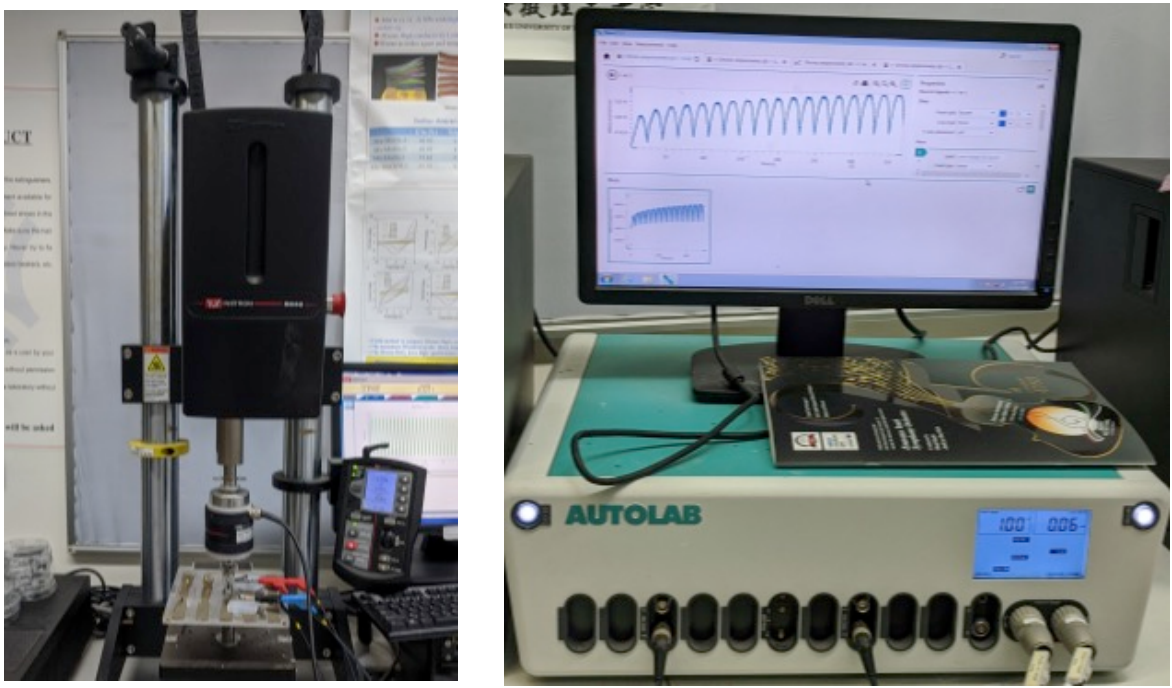

Figure S3 Experimental setup used to obtain the piezo-resistive response.

These graphs shown in Figure S4 (a-e) correspond to Figures 2 & 3 of the main manuscript. Here, instead of compressive stress (MPa), the compressive force (N) is shown in these graphs to show the magnitude of compressive forces used in this work.

The fractional change in resistance (FCR) as a function of force in Newton (N) for loading and unloading regime is shown in Figure S4a. An FCR of ~35% is recorded with an application of a compressive force of ~1800N for the plain cement. For the polymer modified cement, the FCR is ~60% for the same stress. This is due to better transfer of load to the e-textile in the polymer modified cement composite than that of the plain cement. The force sensitivity ($N^{-1}$) can be calculated with the following expression:

Force sensitivity (FS):

$$FS = \frac{FCR}{Compressive\ Load\ (N)}$$

The force sensitivity (FS), estimated using equation (i), for a compressive load of 18 N for plain white cement embedded e-textile is about 0.20 mN$^{-1}$ and for that of polymer-modified e-textile embedded cement is 0.33 mN$^{-1}$. The FCR response of the non-modified e-textile embedded cementitious matrix to applied load is about linear. However, the FCR response of the polymer-modified e-textile embedded cementitious matrix is non-linear due to the presence of a viscoelastic polymer. The FCR response of polymer-modified structure is more prominent due to a more pronounced transfer of load enabled by polymeric chains. The current vs voltage (I/V) curves correspond well with the results of Figure S4a as the change in current with applied load is also more prominent for polymer-modified structure (Figure S4c) than that of the non-modified (Figure S4b). Cyclic FCR responses to different loads (50-500N) are shown in Figure S4d & e for polymer-modified and non-modified structures respectively.

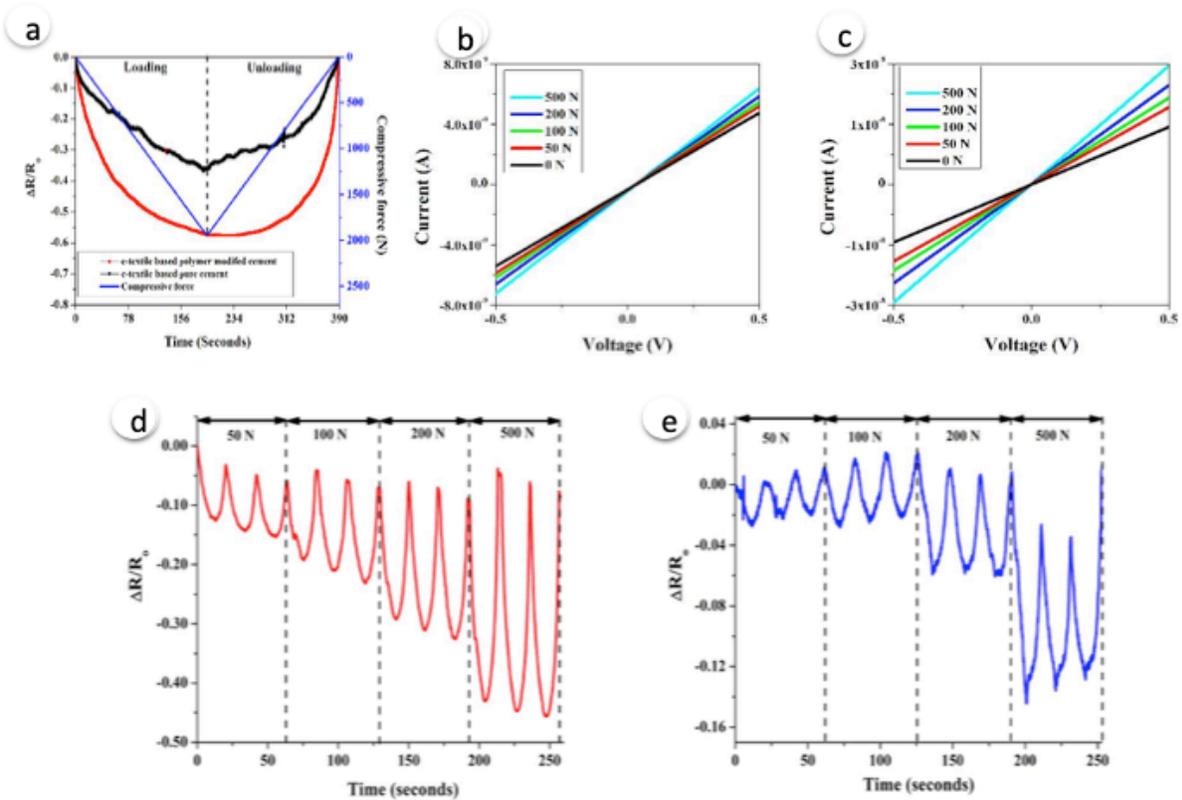

Figure S4 (a) FCR response to direct loading and unloading of e-textile embedded pure cement and polymer-modified cement. (b) IV curves of e-textile embedded pure cement (b) and polymer-modified cement (c) as a function of different cyclic compressive loads. FCR response of e-textile embedded (d) polymer-modified cement (e) and non-modified cement as a function of different cyclic compressive loads.